\documentclass[12pt]{Alvarez}

\usepackage{amssymb}
\usepackage{bm}
\usepackage{subfig}
\usepackage{mathtools}

\begin{document}

\begin{frontmatter}

\title{The importance of stretching rate in achieving true stress relaxation in the elasto-capillary thinning of dilute solutions}

\author[inst1]{Ann Aisling}

\affiliation[inst1]{organization={Department of Chemical and Biological Engineering},
            addressline={Drexel University}, 
            city={Philadelphia},
            postcode={19104}, 
            state={Pennsylvania},
            country={USA, alvarez@drexel.edu}}

\author[inst1]{Renee Saraka}
\author[inst1]{Nicolas J. Alvarez}

\begin{abstract}
This work focuses on inferring the molecular state of the polymer chain required to induce elasto-capillary stress relaxation and the accurate measure of the polymer relaxation time in uniaxial stretching of dilute polymer solutions.
This work is facilitated by the discovery that constant velocity applied at early times leads to initial constant extension rate before reaching the Rayleigh-Plateau instability.
Such constant rate experiments are used to correlate initial stretching kinematics with the thinning dynamics in the elasto-capillary Regime.
We show that there is a minimum initial strain-rate required to induce rate independent elastic effects.
Below the minimum extension rate, insufficient stretching of the chain is observed before capillary instability, such that the polymer stress is comparable to the capillary stress at long times and true stress relaxation is not achieved.
Above the minimum strain-rate, the chain reaches a critical stretch before instability, such that during the unstable filament thinning the polymer stress is significantly larger than the capillary stress and true stress relaxation is observed. 
Using a single relaxation mode Oldroyd-B model, we show that the the minimum strain rate leads to a required initial stretch of the chain before reaching the Rayleigh Plateau limit.
In other words, these results indicate that the chain conformation before entering the Rayleigh Instability Regime determines the elastic behavior of the filament.
Along with the accurate measure of relaxation time, this work introduces a characteristic dimensionless group, called the stretchability factor, that can be used to quantitatively compare different materials based on the overall material deformation/kinematic behavior, not just the relaxation time.
Overall, these results demonstrate a useful methodology to study the stretching of dilute solutions using a constant velocity stretching scheme.
\end{abstract}


\end{frontmatter}

\section{Introduction}
A long standing problem in the literature is the extensional rheological characterization of dilute polymer solutions.
Due to the influence of additional forces, it is very difficult to independently measure the extensional stress of these materials in controlled extensional flows.
Furthermore, the relaxation time of these materials are too fast ($<100$ ms) to actively control the deformation rate.
The most successful methods to date restrict themselves to measuring the relaxation time of the material. 
For example, the capillary breakup rheometer, known as the CaBER, is designed to induce large elastic stresses at early time, such that the relaxation of the material can be measured at later times.
Such methods are better classified as an indexer rather than a rheometer since they do not actively control the deformation or stress.

While such indexing methods have proved useful to determining the relaxation time of known materials, the measurement does not always lead to unique/repeatable results when testing complex unknown fluids.
The cause of this can be argued to arise from poor sample preparation, and/or poorly defined material history \cite{Anna2001}.
However, in this work we show that other reasons can result from inappropriate instrument testing parameters.
Namely we show that the strain on the polymeric chain must be large enough in the initial stretching stages to observe a meaningful relaxation of the system.

The accurate measurement of a material relaxation time requires that the measurement induce a relaxation of the material.
In other words, the stress state of the material must exceed the imposed stress by the instrument at some instant in time, such that the material relaxes as a function of time to a lower stress state.
For details on this type of measurement, see for example ``stress relaxation after shear flow'' in Dynamics of Polymeric Liquids.\cite{DPLvolume1}
In stress relaxation experiments, the material relaxation rate is determined by either keeping the stress constant or the strain constant and measuring the decay rate of the other. 
All indexing methods mentioned above work on measuring the decay of strain with a relatively constant imposed stress, e.g. capillary stress.
However, it should be clearly understood that the stress state is neither controlled nor exactly constant in such instruments.

Since many of these apparatuses have different acronyms and variations in methodologies, it may not be obvious to the reader that almost all of them work on the same filament thinning principles, e.g. CaBER \cite{campo-deano2010,Rodd2005,Arnolds2010,Yesilata2006,Oliveira2006}, drop on a substrate (DoS)\cite{Amarouchene2001,Cooperwhite2002,Dinic2017,Dinic2019,Hsiao2017,Jimenez2018,Rosello2019,DeBlais2020a}, RoJER \cite{Mathues2018,Sharma2015,Keshavarz2015}, and pulsed surface acoustic wave on a drop [SAW]\cite{Bhattacharjee2011}.
All these extensional relaxation techniques, utilize uniaxial deformation to induce extensional stress and to monitor the relaxation process. 
Initially the stress state of the material is imposed (not controlled) by the rate of motion of two fractions of the material stretched away from each other, henceforth referred to as \textit{active stretching}.
At some prescribed time, the bulk stretching is stopped and the stress state imposed on the material is due to interfacial stresses, henceforth referred to as \textit{passive stretching}.
For a molecular relaxation process to be observed, the stress state of the material must be higher than the interfacial stress at the transition from active stretching to passive stretching .
However, as noted above the stress state of the material is not controlled and therefore this requirement is not guaranteed, and, as we will show, is rarely met when active stretching is terminated.
Moreover, the interfacial stress of the material is increasing with increasing strain, which makes this a non-ideal measurement.
Overall, this type of measurement does not necessarily guarantee or lead to a clear measurement of one or more relaxation times of a given material.

In this work, we use a uniaxial extensional rheometer in constant velocity mode to demonstrate that there is a minimum extension rate and strain of a model polyethylene oxide solution that guarantees accurate measurement of the material relaxation time. 
When these conditions are not met, an erroneous relaxation time can be extracted from the data, but it significantly under predicts the true relaxation time of the material, and is a function of initial active stretching rate.
Furthermore, using the classical theory of Hinch and coworkers, we show that a higher material stress than interfacial stress, requires a minimum chain stretch at the transition from active to passive stretching.
Thus, by coupling experiments of changing initial active stretching rate with theory, one is able to better understand the molecular stretch of the material that induces strong elastic effects
Lastly, we introduce an additional parameter, referred to as the stretchability factor, that relates the overall behavior of the fluid with the applied kinematics.
This new easy to measure parameter quantifies the geometric stretching of the material in the context of the applied kinematics, which is typically done using complex image analysis.

\section{Materials and Methods}

\begin{figure}[h!]
    \centering
    \includegraphics[width=0.95\textwidth]{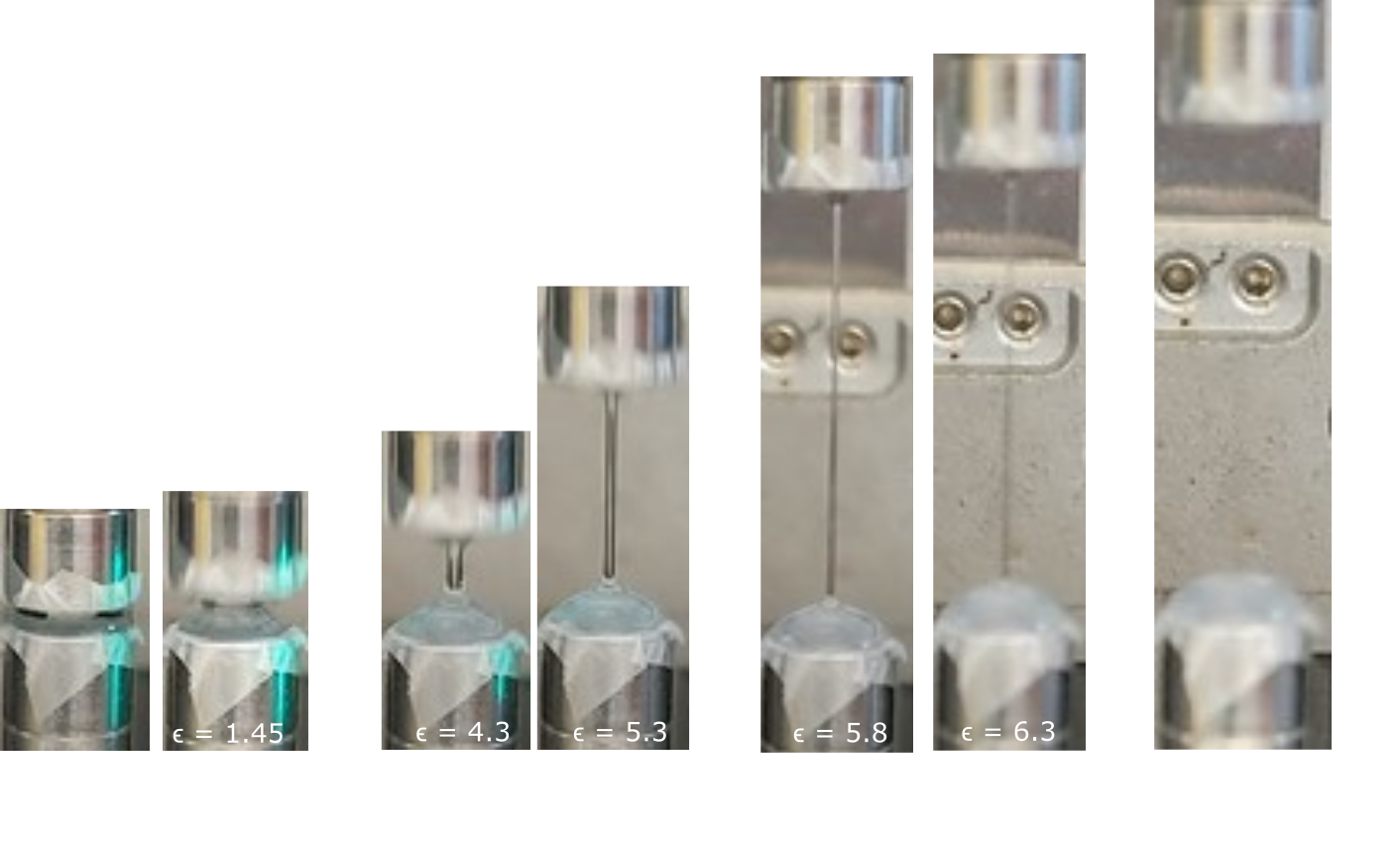}
    \caption{Capillary necking and filament stretching of a dilute polymer solution on the VADER } \label{fig:schematic}
\end{figure}

Solutions of polyethylene oxide (Sigma Aldrich) with an average molecular weight of 4$\times10^6$ g/mol in water were prepared in concentrations of 10, 30, 60, 120, 240, 480, and 960 mg/L (wppm) by dissolving PEO in water, stirred, and allowed to sit for at least 24 hours to ensure homogeneity.\cite{DeBlais2020a}
Note that the overlap concentration for this Mw PEO has been previously measured to be 770 wppm.\cite{Graessley1980}
The dissolved solutions were tested on an 8 mm parallel plate geometry attached to a Versatile Accurate Deformation Extensional Rheometer (VADER – Rheo Filament) at constant velocity separation of the top plate.
The VADER software was changed to allow for a higher data collection rate (from 33 to 100 points/s) for these experiments.
The maximum measured Hencky strain for the VADER is 8.76 based on the reported minimum measured diameter of 100 $\mu$m.
The stainless-steel plates were covered with tape (adhesive side not in contact with sample) to avoid sample slipping. 
Samples were loaded at constant volume (100 µL) onto the parallel plates using a micropipette. 
The top plate was lowered so that the diameter of the sample was approximately equal to that of the geometry plates (between 7.6-7.9 mm). 
The sample was scanned with the intrinsic laser micrometer to measure the initial diameter and shape. 
Uniform sample shape was visually confirmed by the user to ensure that no premature stretching occurred, and that the sample diameter was axisymmetric. 
The sample was allowed to relax in this state for a minimum of 30 seconds before stretching. 
The sample was stretched at constant velocity between 0.1-100 mm/s. 
Each concentration and velocity were repeated using three fresh sample loadings to ensure reproducibility.
Furthermore each sample was tested three consecutive times (with sufficient time >1 min for the sample to fully relax between runs) and the average of these measurements are reported. 
Two different researchers ran sample sets in an initial gauge analysis to identify, quantify, and manage sources of experimental variability and error - such as the length of time samples needed to relax, slip of material on plate, impact of bubbles, evaporation, and age of prepared solution (all of these matter).  
Minimum sample diameter and separation height of plates were collected as a function of time and analyzed to calculate strains and model stresses. 
Table \ref{tab:variables} shows a list of all variables used to analyze and discuss the experimental data.  
Note that some of these are common in the extensional rheology literature, however we have introduced several new variables such as stretchability, $S$ and stretchability plateau.
Figure \ref{fig:schematic} shows a time series of images before and after applying a constant velocity.

\begin{table}[h!]
\begin{tabular}{ll}
\hline
\multicolumn{1}{|l|}{Term} & \multicolumn{1}{l|}{Equation}                                                                                                                                                   \\ \hline
\multicolumn{1}{|l|}{Radial Hencky strain, $\epsilon_D$}                  & \multicolumn{1}{l|}{$\epsilon_D=2\ln\bigg(\frac{D_0}{D(t)}\bigg)$}                                                                                                             \\ \hline
\multicolumn{1}{|l|}{Axial Hencky strain, $\epsilon_z$}                   & \multicolumn{1}{l|}{$\epsilon_z=\ln\bigg(\frac{H}{H_o}\bigg)$}                                                                                                                  \\ \hline
\multicolumn{1}{|l|}{Local Strain Rate, $\dot{\epsilon}_{D,\rm{local}}$}  & \multicolumn{1}{l|}{$\dot{\epsilon}_{D,\rm{local}}=\frac{\epsilon_D(t_{i+2})-\epsilon_D(t_i}{2\Delta t}$}                                                                             \\ \hline
\multicolumn{1}{|l|}{Steady State Strain Rate}                            & \multicolumn{1}{l|}{$\dot{\epsilon}_D=\frac{d\epsilon}{dt}=\frac{\sum_{i=1}^N\epsilon_i t_i-\sum_{i=1}^N t_i\sum_{i=1}^N\epsilon_i}{N\sum_{i=1}^N(t_i^2)-(\sum_{i=1}^Nt_i)^2}$} \\ \hline
\multicolumn{1}{|l|}{Relaxation Time}                                     & \multicolumn{1}{l|}{$\lambda=\frac{2}{(3\dot{\epsilon}_D)}$}                                                                                                                         \\ \hline
\multicolumn{1}{|l|}{Stretchability, $S$}                      & \multicolumn{1}{l|}{$S=\bigg(\frac{\partial\epsilon_Z}{\partial\epsilon_D}\bigg)$}                                                                                                             \\ \hline
\multicolumn{1}{|l|}{Stretchability Plateau}                                      & \multicolumn{1}{l|}{$\lim_{t\to\infty}S$}                                                                                                                                       \\ \hline
\end{tabular}
\caption{List of variables used in the analysis of constant velocity extensional data}\label{tab:variables}
\end{table}

\section{Single Relaxation Oldroyd-B Model}

Direct measure of the chain stretch is only possible in a specialized experiments, such as in the case of labeled DNA. \cite{Schroeder2003}
Therefore, we use a finite extensibility nonlinear elastic (FENE) model, developed by Entov and Hinch \cite{Entov1997}, to estimate the chain configuration and polymer stress as a function of strain using experimental data. 
This model has also been used by others to better understand filament thinning experiments.\cite{Wagner2015,Mathues2018}
We quickly detail the equations solved to estimate the chain stretch, $A_{zz}$, and the polymer stress, $\sigma_p$.
Consider a uniform cylindrical liquid bridge whose radius, $R(t)$, is decreasing as a function of time. 
Let the axial strain-rate of the axisymmetirc extensional flow be $\dot{\epsilon}(t)$, such that the radius decreases via:
\begin{equation}
    \frac{\partial R}{\partial t}=-\frac{1}{2}\dot{\epsilon}R(t)
\end{equation}
from the definition of the Hencky strain, $\epsilon=-2\ln(D(t)/D_0)$, and the Hencky strain rate, $\dot{\epsilon}=\partial \epsilon/\partial t$.
Taking the constituitive model to be the Oldroyd-B fluid with one relaxation mode \cite{Frigaard2022}, which is equivalent to the FENE model assuming a single relaxation mode (i.e. the longest relaxation mode).
The elastic deformation of the polymer coil is described by the normalized conformation tensor, $\bf{A}$, i.e. a fully relaxed chain has $A_{zz}=A_{rr}=1$ and the fully stretched chain is given by $A_{zz}=A_{rr}=L^2$.
Taking into account the microstructural evolution equation:
\begin{equation}
    \overset{\nabla}{\bf{A}}=\frac{D \bf{A}}{Dt}-\bf{A}\nabla\bf{v}-(\nabla\bf{v})^T\bf{A}=-\frac{1}{\lambda}\bf{\sigma_p}
\end{equation}
where $\overset{\nabla}{\bf{A}}$ denotes the upper-convected derivative, $\lambda$ denotes the relaxation time of the polymer chain, and $\bf{\sigma_p}$ is the polymeric stress given by
\begin{equation}
    \bf{\sigma_p}=G\epsilon_{chain}=G(Z\bf{A}-\bf{I})
\end{equation}
where $G$ is the shear modulus, the parameter $Z$ denotes the correction term accounting for the nonlinearity and the finite extensibility of the chain of fully stretched length $L$, given by,
\begin{equation}
    Z=\frac{L^2}{L^2-tr(\bf{A})}=\frac{L^2}{L^2+3-A_{zz}-2A_{rr}}
\end{equation}
where $tr(\bf{A})$ denotes the trace of the conformation tensor.
From the radial symmetry and uniaxial flow kinematics the evolution equations reduces to two differential equations for $A_{zz}$ and $A_{rr}$ respectively:
\begin{align}
        \frac{d A_{zz}}{d t}&=2\dot{\epsilon}A_{zz}-\frac{Z}{\lambda}(A_{zz}-1)\\
        \frac{d A_{rr}}{d t}&=-\dot{\epsilon}A_{rr}-\frac{Z}{\lambda}(A_{rr}-1)
\end{align}
These equations can be solved subject to the initial conditions of the chains in there fully relaxed state.
\begin{equation}
    A_{zz}(0)=A_{rr}(0)=1
\end{equation}
The only parameters that need to be specified are $L$, $\lambda$, and $\dot{\epsilon}$.
For 4$\times10^6$ kg/mol PEO, $L=227$ determined using Ref. \cite{Mathues2018, Brandrup1999}
$\lambda$ is determined experimentally as detailed below.
\begin{table}[h!]
\caption{Parameters used in the calculation of L}
\begin{tabular}{llll}
\textbf{Parameter}       & \textbf{Description  }                                                                                                      & \textbf{Value}    & \textbf{Source}                          \\ \hline
$\Theta_B$      & Average Bond Angle                                                                                                 & 1.909    & \cite{Brandrup1999}             \\
j               & Number of Bonds of a Monomer                                                                                       & 3        & \cite{Brandrup1999}             \\
$Mw_0$           & [kg/mol] of monomer                                                                                                  & 0.044    & \cite{Elbing2021,Grandbois1999} \\
$C_{\infty}$ & ratio for a given monomer                                                                                          & 44.1     & \cite{Brandrup1999}             \\
$\nu$           & excluded volume coefﬁcient  & 0.544    & \cite{Clasen2006}               \\
$\mu$           & [Pa s] viscosity of solvent                                                                                        & 0.001    & \cite{Brandrup1999}              \\
$\rho$          & [kg/m$^3$] density                                                                                                     & 998      &                                 \\
$\gamma$        & [N/m] surface tension                                                                                                & 0.062    & \cite{Kim1997}                  \\
$R_0$             & [m] radius of plates                                                                                                 & 0.004    &                                 \\
Mw             & [kg/mol] of polymer chain                                                                                            & 4$\times10^3$ &    \\   \hline                             
\end{tabular}
\end{table}
We depart from the typical solution method, which is to estimate the strain-rate from a thin filament model \cite{Wagner2015}, and instead use experimental data of $D(t)$ to determine the local experimental strain-rate, $\dot{\epsilon}$ as a function of time as an input into the equations.
The experimental $\dot{\epsilon}$ values are smoothed using a Whitaker smoothing routine and the values are used directly in the calculation of chain stretch and polymer stress. 
The equations are solved using matlab ODE15s and the script file is available upon request.
The estimated maximum coil length of the polymer, the finite extensibility parameter, is determined by the molecular weight and the excluded volume coefficient characterizing the quality of the solvent.\cite{Wagner2015, Clasen2006}

\section{Results} 

There are typically three sequential regimes of filament stretching for a dilute polymer solution: (1) visco-capillary, (2) Rayleigh instability Regime, and (3) elasto-capillary.
Regimes (1) and (3) are defined by the magnitude of the individual force components, i.e. capillary force, viscous force, and elastic force.
For example, the visco-capillary thinning Regime is a competition of shear viscosity and capillary forces, while Regime (3) is a competition between capillary and polymer elastic forces.
The instability Regime, referred to by Entov and Hinch \cite{Entov1997} as the middle-elastic time, is defined by an instability caused by capillary forces and is commonly referred to as the Rayleigh-Plateau instability.
This instability occurs at a critical aspect ratio depending on the volume of the filament.\cite{Barakat2021}
Note that a Newtonian fluid, with no elastic effects, is only capable of Regime (1) and (2).\cite{Eggers2008}
Polymer solutions stretched fast enough to mount sufficient stored elastic energy during Regime (1) can skip Regime (2) and immediately observe Regime (3), as in the experiments of Liang and Mackley.\cite{Liang1994}
Otherwise, Regime (3) is achieved by very fast stretching during Regime (2) such that the elastic stress is comparable or exceeds the capillary stress.\cite{Amarouchene2001,Goldin1969,Entov1984}
If Regime (3) can be achieved such that the elastic stress exceeds the capillary stress, the chain will undergo stress relaxation.
Thus, Regime (3) is typically analyzed to determine a polymer relaxation time from the thinning filament. \cite{Rodd2005,Anna2001,Clasen2006,Bazilevskii1997}

\begin{figure}[h!]
    \centering
    \includegraphics[width=0.95\textwidth]{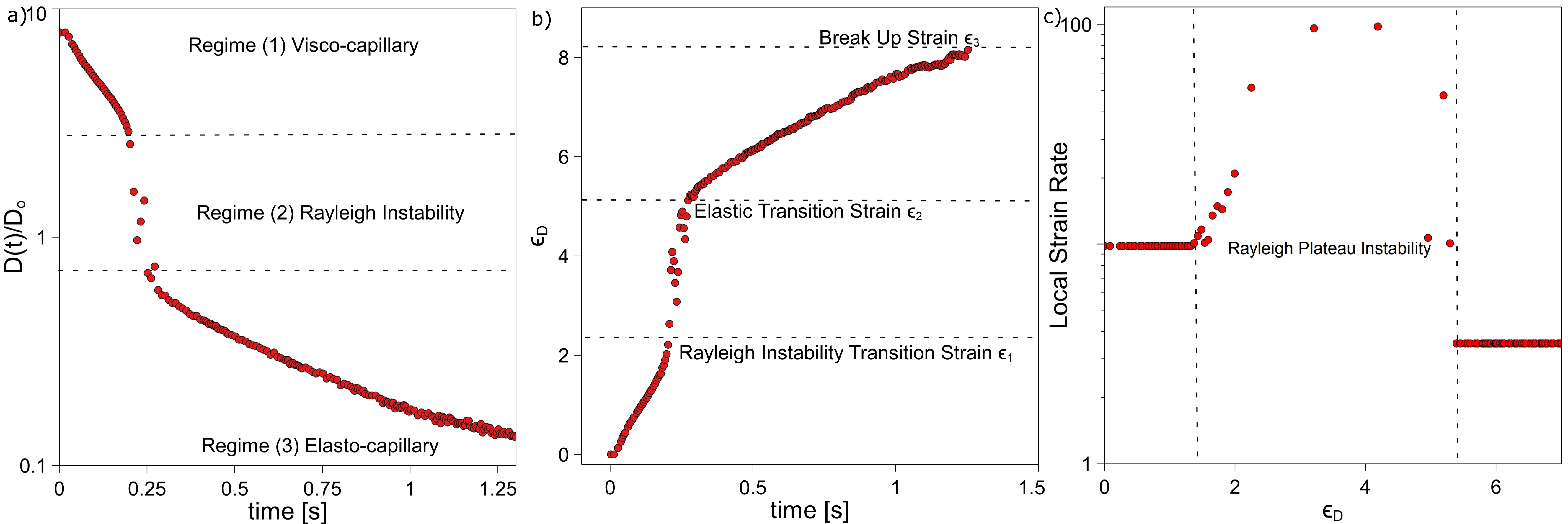}
    \caption{(a) 960 wppm 20 mm/s Normalized Diameter Decay vs time, (b) 960 wppm 20 mm s$^{-1}$ $\epsilon_D$ vs time, (c) 960 wppm 7 mm s$^{-1}$ Local Strain Rate vs Strain } \label{fig:960Local}
\end{figure}

In the case of low-viscosity dilute solutions, it is rare that Regime (2) is avoided.
Therefore, it is not necessarily the case that the elastic stresses are larger than the capillary stresses in Regime (3).
The three Regimes are typically visualized by monitoring the normalized minimum diameter of a liquid filament as a function of time.
Figure \ref{fig:960Local}(a) shows the initial Newtonian visco-capillary necking Regime, followed by the Rayleigh Plateau instability transition, and finally by the elastic stretching Regime and final break-up. \cite{Christanti2001,Bousfield1986}
Note that each Regime has a different decay of diameter with time.
We hypothesize that the degree to which the steady elastic Regime is observed, strongly depends on the state of the chain upon entering Regime (2), i.e. the stretch of the polymer chain in Regime (1) dictates the elastic effects. 
Thus, this work determines the minimum strain-rate in Regime (1) required to observe steady elastic stretching in Regime (3), and shows that this critical rate is related to the relaxation time of the polymer chain.

For the purpose of this paper, we are interested in the magnitude of the radial Hencky strain, $\epsilon_D$ as a function of time.
Figure \ref{fig:960Local} (b) shows Hencky strain versus time for the data in Figure \ref{fig:960Local}(a).
Furthermore, the rate of change of radial Hencky Strain, known as the Hencky strain rate, $\dot{\epsilon}_D$, is proportional to the inverse of the relaxation time \cite{McKinley2000,Mckinley2005,Dinic2019,Snoeijer2020}.
Figure \ref{fig:960Local} (c) shows the local Hencky strain rate as a function of time, which will be later used to calculate other derivative parameters, such as coil stretch and relaxation time. 
The three stretching Regimes are clearly marked in Figure \ref{fig:960Local}.
In Figure \ref{fig:960Local}(b) we observe that for this data each Regime occurs over approximately two units of strain; however, the rates in these Regimes are very different.
It is useful to define three characteristic strains: $\epsilon_1$, the Rayleigh instability transition strain, $\epsilon_2$, the elastic transition strain, and $\epsilon_3$, the maximum measured strain.
Note that $\epsilon_3$ is a measure of the instrument limitation in accurately measuring the diameter of the filament ($\epsilon_3\approx8$ for the VADER).
The rates of deformation observed in Figure \ref{fig:960Local}(c) show a very interesting result regarding the constant velocity scheme: namely that the constant velocity scheme results in a constant rate of extension in Regime (1).
Regime (2) shows an unstable stretching rate that grows to a maximum\cite{Sattler2012}, and then decreases to a steady state value in Regime (3).
The steady state value of strain-rate observed in Regime (3) is what previous studies typically use to calculate the relaxation time of the material \cite{Turkoz2018}.
We will demonstrate that the strain rate in Regime (1) must be equal to or larger that the rate observed in Regime (3) to accurately measure the longest relaxation time.

 \begin{figure}[h!]
    \centering
    \includegraphics [width=0.95\textwidth]{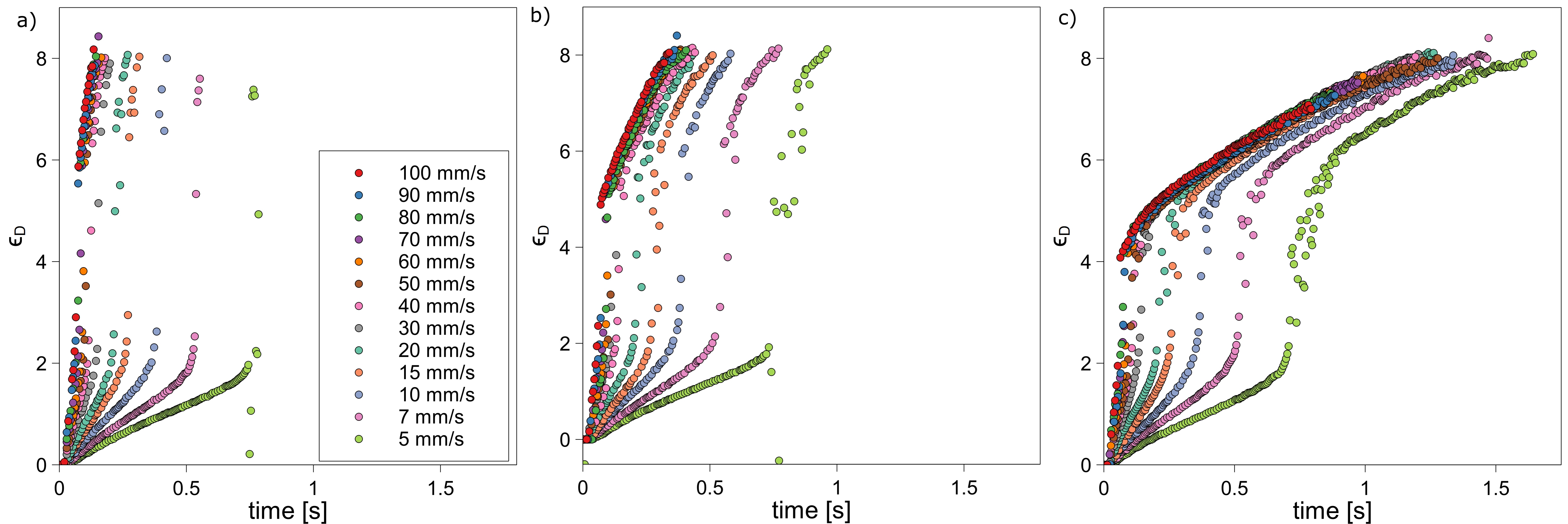}
    \caption{Hencky Strain vs time for different velocities for three concentrations a) 30 wppm, b) 240 wppm, and c) 960 wppm}
    \label{fig:edvt}
\end{figure}

Figure \ref{fig:edvt} shows Hencky strain as a function of time for all velocities and three concentrations measured in this study.
For all concentrations, we observe that Regime (1) is shortening in time for increasing velocity.
In other words, the rate of deformation in Regime (1) is increasing with increasing velocity.
Furthermore, we observe $\epsilon_1$ is independent of the velocity and the concentration and that $\epsilon_2$ is decreasing with increasing velocity for a given concentration.
Lastly, we observe that the rate of extension in Regime (2) and Regime (3) are a function of velocity and polymer concentration.

\begin{figure}[h!]
    \centering
    \includegraphics[width=0.5\textwidth]{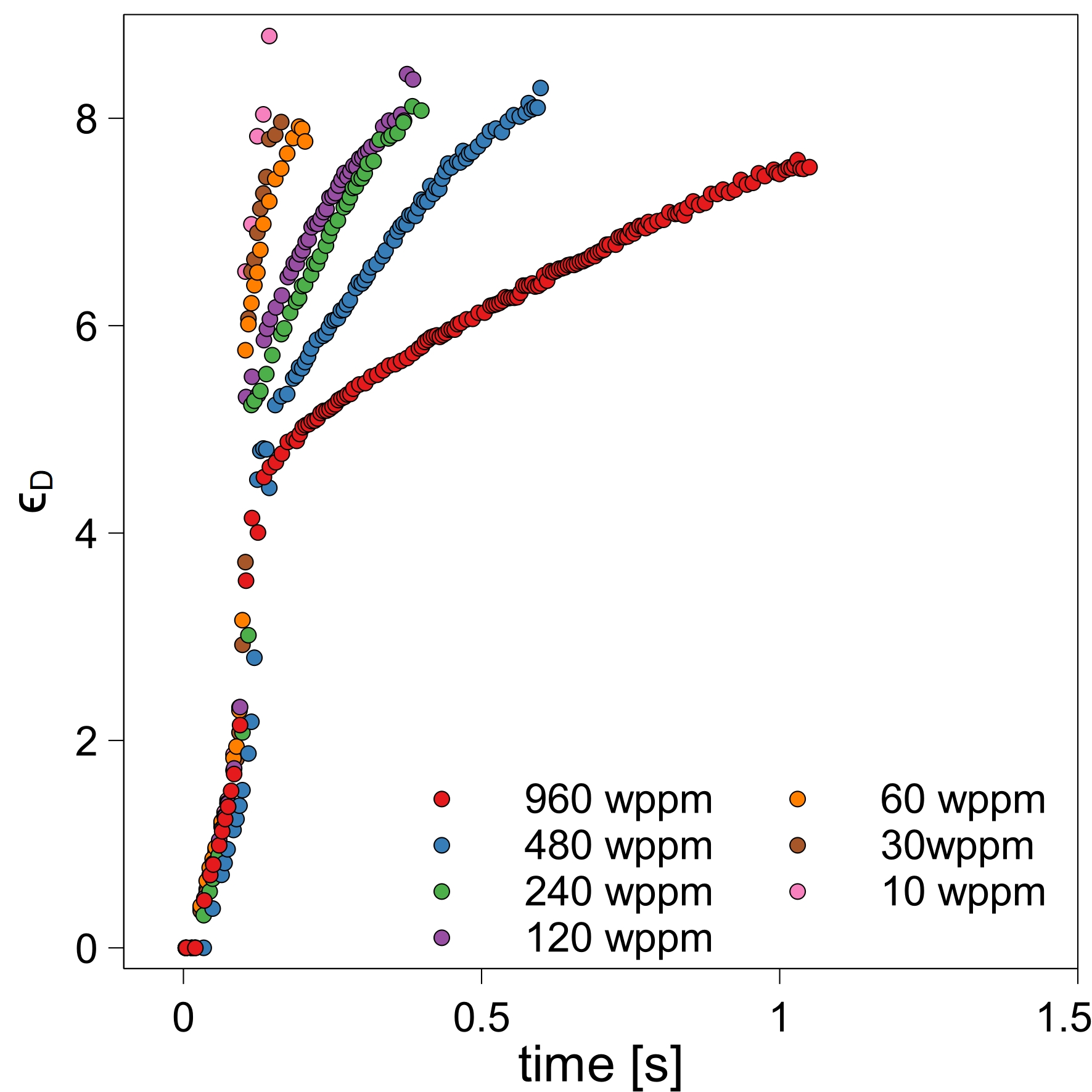}
    \caption{Various Concentrations of Polymer in Solution Measured at 50 mm/s}
    \label{fig:evstConcentration}
\end{figure}

Figure \ref{fig:evstConcentration} shows Hencky strain versus time for seven measured concentrations stretched at the same velocity.
This figure helps to discern which features of Figure \ref{fig:edvt} are rate and concentration dependent.
For example, we observe that all seven concentrations have the same initial extension rate in Regime (1), indicating that the initial rate is simply a function of the applied velocity.
Conversely, we observe that the transition strain from Regime (2) to Regime (3), i.e. $\epsilon_2$, is a function of polymer concentration for the same applied initial rate.
Lastly, the extension rate observed in Regime (3) is a strong function of polymer concentration, indicating that this rate is a material parameter.
As one might expect, the lowest and highest concentration have the highest and lowest strain-rate in Regime (3), respectively, which tracks with the inverse of their expected relaxation times.
These important observations are discussed in detail below, with an emphasis on the measurement of relaxation time.

\section{Discussion}

\begin{figure}[h!]
    \centering
    \includegraphics [width=0.95\textwidth]{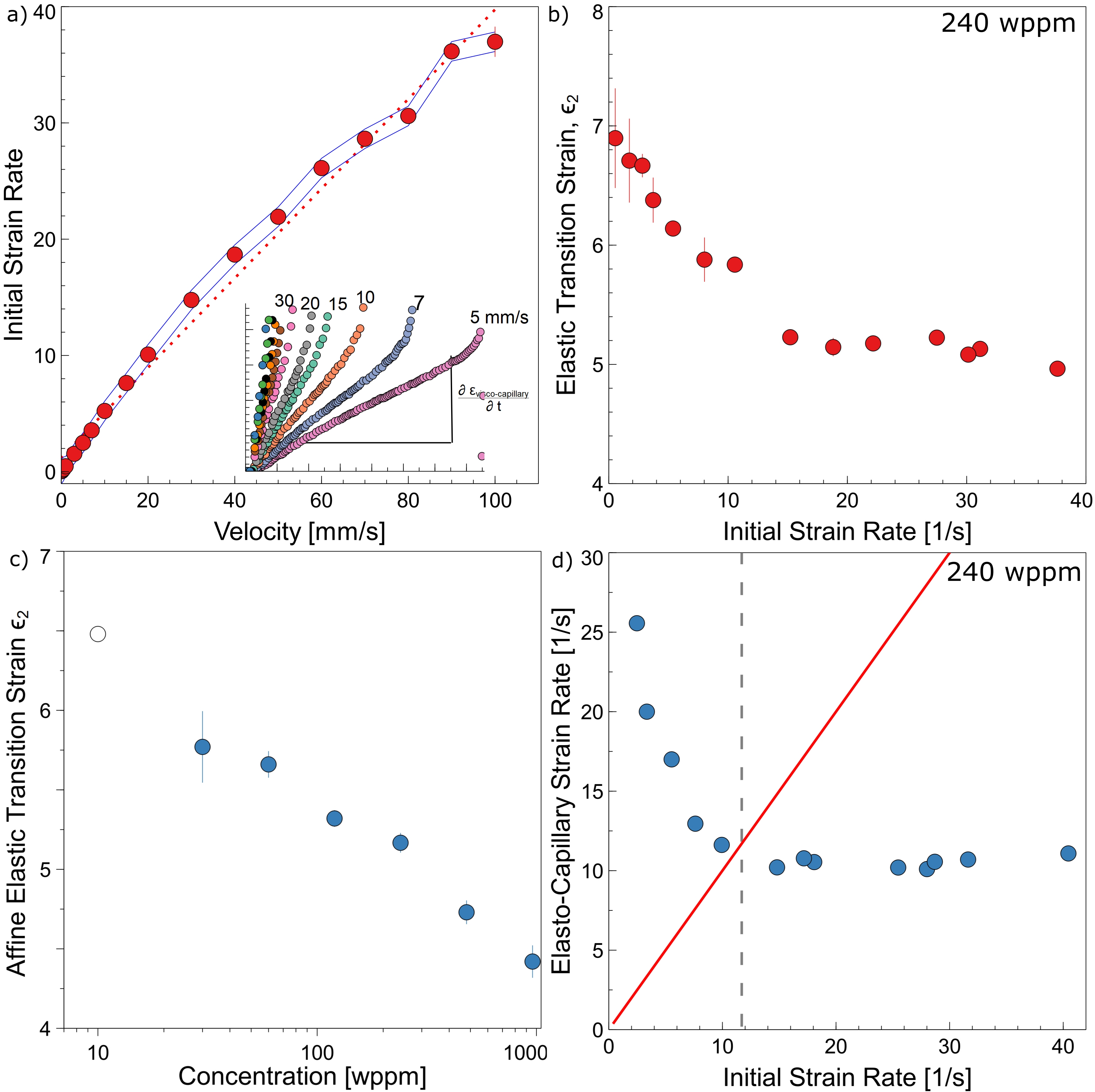}
    \caption{(a) initial measured strain rate as a function of stretching velocity, inset shows raw strain data (b) $\epsilon_2$ as a function of initial strain rate for 240 wppm, (c) asymptotic $\epsilon_2$ as a function of concentration, and (d) elasto-capillary thinning rate as a function of initial strain rate for 240 wppm.}    \label{fig:edvtAnalysis}
\end{figure}

One surprising result from Figure \ref{fig:edvt} is that the constant velocity measurement results in a constant rate experiment in Regime (1).
This is better seen in Figure \ref{fig:edvtAnalysis}(a) where the average strain-rate of Regime (1) is plotted as a function of applied velocity.
Recall that, as shown in Figure \ref{fig:evstConcentration}, the strain-rate in Regime (1) is not a function of polymer concentration, but is a property of the sample aspect ratio and solution viscosity, which does not vary with concentration for dilute solutions. 
Note that increasing sample aspect ratio leads to a lower initial extension rate for the same velocity, and vice versa.
Different plate sizes, and therefore aspect ratios, would result in a different linear relationship with velocity, and different minimum velocities.
Figure \ref{fig:edvtAnalysis}(a) shows a linear dependence of the initial strain rate on the plate separation velocity, with a slope of approximately 0.5.
The confidence interval of the population varies narrowly with a standard error that increases linearly with velocity, up to a value of 1.4 at 100 mm/s. 
The fact that the constant velocity experiment leads to a constant rate is advantageous since polymer physics is formulated in terms of normalized timescales, such as the Weissenberg number \cite{Poole2012}, that involve the rate of extension compared to the polymer relaxation time.
Thus the discussion of experimental data will be done in terms of initial extension rate not velocity. 

Another observed trend that is important to highlight is the effect of initial strain rate and concentration on the elastic transition strain, $\epsilon_2$, described in Figures \ref{fig:960Local} and \ref{fig:edvt}.
Figure \ref{fig:edvtAnalysis} (b) shows $\epsilon_2$ as a function of initial strain rate for $C=240$ wppm.
We clearly observe a decreasing function of $\epsilon_2$ for increasing initial strain-rate below 12/s. 
After which, we observe a plateau in $\epsilon_2$, demonstrating a rate independent phenomenon.
One might propose that this transition in $\epsilon_2$ is related to a transition of stretching the chains at a rate below the inverse relaxation time to stretching faster than the inverse relaxation time.  
In other words, the strain-rate where $\epsilon_2$ becomes rate-independent is a direct measure of the inverse of the polymer relaxation time.
It is at this initial strain-rate that we would expect the polymer stress to exceed the capillary stress, allowing for a direct observation of the polymer relaxation time in the elasto-capillary regime.
Moreover, the fact that $\epsilon_2$ is rate-independent above a minimum initial strain-rate suggests affine deformation of the polymer molecules above this critical rate.

Figure \ref{fig:edvtAnalysis}(c) shows the asymptotic $\epsilon_2$ as a function of concentration, also tabulated in Table \ref{tab:Results}.
As the concentration increases the asymptotic $\epsilon_2$ decreases.
This suggests that there is a critical stress for a given initial sample aspect ratio required to achieve true stress relaxation.
Current polymer physics scaling laws establish that relaxation time and polymer stress increase with increasing polymer concentration.
More specifically, polymer stress, $\sigma_p$ depends directly on the solution viscosity (for dilute solutions) and shear modulus, $G$, (for concentrated solutions) which are explicit functions of polymer concentration \cite{Entov1997}.
This implies that a higher concentration requires less strain (i.e. chain stretch) to achieve a given $\sigma_p$, which is the trend observed in Figure \ref{fig:edvtAnalysis}(c).
In other words, lower concentrations solutions require significantly higher strains than concentrated solutions to observe stress relaxation.  
Thus, an instruments accessible strain window will limit the measurable concentration for a given polymer.

The practical importance of the elastic transition strain and its relation to the relaxation time are made evident in Figure \ref{fig:edvtAnalysis} (d).
We observe that the elasto-capillary strain-rate measured in Regime (3) is initially a decreasing function of the initial strain rate, until it reaches an asymptotic value.
The solid line in Figure \ref{fig:edvtAnalysis} (d) is the $y=x$ line, which crosses the experimental data at the asymptotic elasto-capillary strain rate.
This phenomena occurs in each concentration tested and reveals a fundamental kinematic requirement of measurement: the initial rate must be equal to or faster than the inverse relaxation time of the polymer solution in order to observe the correct timescale.
In other words, the initial extension rate must be large enough that the initial Weissenberg number, $\mathrm{Wi}=\lambda\dot{\epsilon}>2/3$, such that the chains are undergoing stretching in the visco-capillary regime.
When this criteria is met, the chains are undergoing stress relaxation in the elasto-capillary regime and the measured relaxation time is therefore a material parameter, see Table \ref{tab:Results}
Most importantly, an observed extension rate in the elasto-capillary regime does not necessarily correspond to the materials relaxation time, and any attempt to extract a relaxation time below $Wi<2/3$ will lead to erroneously smaller relaxation times.
This occurs most likely because $\sigma_p$ is not sufficiently greater than the capillary stress to observe true stress relaxation.
This is true for all concentrations measured.

Table \ref{tab:Results} shows all the measured and derived values for the seven concentrations studied.
As noted in the table, the lowest initial strain-rate (denoted as initial $\dot{\epsilon}$) where an asymptotic elasto-capillary thinning rate is observed is either larger or approximately equal to the measured Elastic $\dot{\epsilon}$.
Above this initial rate, the Elastic $\dot{\epsilon}$ is independent of the initial $\dot{\epsilon}$.
The asymptotic relaxation time for each concentration was calculated using the definition in Table \ref{tab:variables}.
One exception is 10 wppm, whereby no asymptotic elastic $\dot{\epsilon}$ was observed for the achievable rates using the VADER.
This brings up the important point that not all relaxation times are accessible by a machine with a limited initial plate separation velocity range.
For example, using the relationship between initial $\dot{\epsilon}$ and plate velocity in Figure \ref{fig:edvtAnalysis} for the sample aspect ratio used in this study, a predicted minimum plate separation velocity, $V_{min}$ required to measure a relaxation time, $\lambda$, is given by
\begin{equation}
    V_{min}=\frac{890 [mm]}{\lambda [ms]}
\end{equation}
This equation allows the limit of a given experimental setup to be determined simply by knowing its maximum velocity.

However, initial plate separation velocity is not the only limitation to accurately measuring material relaxation times.
Another important limit is the issue of limited data collection speeds.
For example, although the VADER can achieve velocities up to 600 mm/s, which would theoretically allow for the accurate measurement of $\lambda=1.45$ ms, it is limited by data a collection rate of 50 points per second, which limits the VADER to approximately 40 1/s.

\begin{table}[h!]
\caption{Asymptotic values of derived variables from constant velocity stretching of different PEO concentrations.  Note that the Elastic $\dot{\epsilon}$ represents an average value and that the * indicates the minimum measured derived variable not the asymptotic value, as no asymptote was achieved.} \label{tab:Results}
\begin{tabular}{|c|c|c|c|c|c|}
\hline
\multicolumn{1}{|l|}{\textbf{C [wppm]}} & \multicolumn{1}{l|}{\textbf{C/C$^*$}} & \multicolumn{1}{l|}{\textbf{Initial $\dot{\epsilon}$}} & \multicolumn{1}{l|}{\textbf{Elastic $\dot{\epsilon}$ [1/s]}} & \multicolumn{1}{l|}{\textbf{Relaxation Time [ms]}} & \multicolumn{1}{l|}{\textbf{$\epsilon_2$}} \\ \hline
10                                      & 0.013                                 & 40*                                                    & 67.1                                                         & 9.94*                                              & 6.48*                                       \\
30                                      & 0.039                                 & 43.8                                                   & 33.5                                                         & 19.92                                              & 5.77                                       \\
60                                      & 0.078                                 & 24.3                                                   & 21.6                                                         & 30.88                                              & 5.66                                       \\
120                                     & 0.156                                 & 11                                                     & 11.0                                                         & 60.64                                              & 5.32                                       \\
240                                     & 0.312                                 & 11.7                                                   & 10.5                                                         & 63.36                                              & 5.167                                       \\
480                                     & 0.623                                 & 8.1                                                    & 7.31                                                         & 91.26                                              & 4.73                                       \\
960                                     & 1.247                                 & 3.35                                                   & 3.36                                                         & 198.64                                             & 4.42              \\                        \hline
\end{tabular}
\end{table}

\begin{figure}[h!] 
    \centering
    \includegraphics[width=0.95\textwidth]{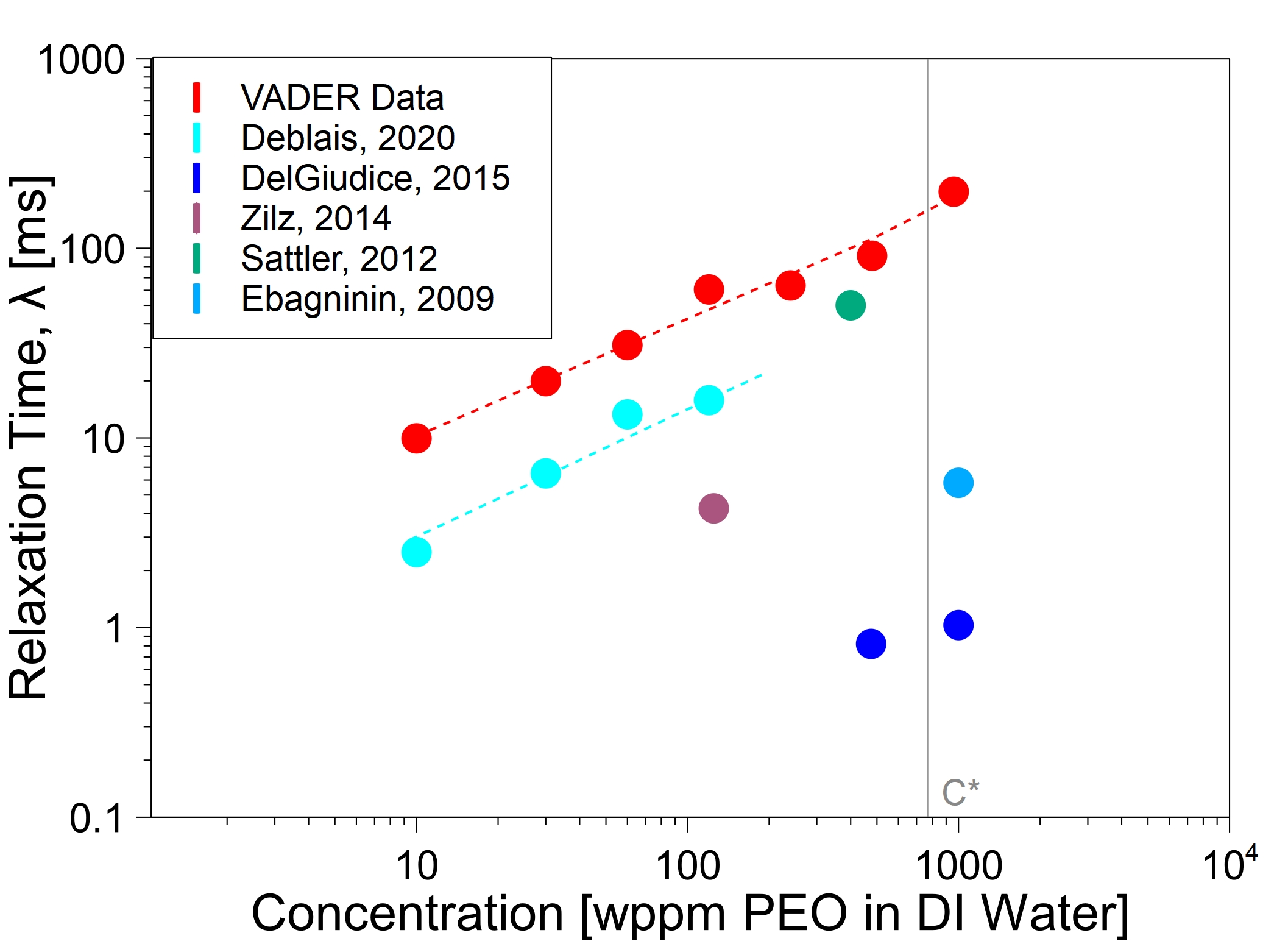}
     \caption{Comparison of relaxtion time measured in this study compared to relaxation times reported in the literature for the same molecular weight PEO \cite{DeBlais2020a, DelGiudice2015,Zilz2014,Sattler2012,Ebagninin2009}}     \label{fig:Relaxation Time}
\end{figure}

The asymptotic relaxation times in Table \ref{tab:Results} are compared with previously published results in Figure \ref{fig:Relaxation Time} for the same molecular weight PEO. 
Note that the data for this work and that of Deblais et al. use identical material sourcing and target concentrations for direct comparisons.
Although the trends in $\lambda$ with concentration are similar between the two studies, the values measured using DOS are consistently lower. 
We denote that the difference is related to the inability for DOS to achieve the required initial extension rate to measure the correct relaxation time.
Surprisingly, all the previously reported results for this molecular weight PEO are lower than the asymptotic relaxation time measured in this study, suggesting that previous studies did not achieve the necessary initial extension rate to measure stress relaxation.
Another explanation of the discrepancy in the data is that even if the initial rate was achieved, some techniques only apply a finite strain to the filament before allowing the filament to undergo unstable thinning.
It is possible that the finite strain may not be large enough to ensure that the polymer stress is larger than the capillary stress, and elasto-capillary stress relaxation is achieved. 

The results and discussion above suggest that the state of the polymer chain just before entering the instability is a key parameter to measuring stress relaxation in the elasto-capillary regime.
To test this idea, we used a single relaxation Oldroyd-B model to estimate the chain stretch as a function of strain during different stretching velocities.
The experimental $D_{min}(t)$ was used to determine $\dot{\epsilon}(t)$ and used as an input into the model to calculate $A_{zz}$ and $\sigma_p$ as a function time and filament strain, see Model section for details. 
A subset of the experimental data used for the calculation are in Figure \ref{fig:240wppmAzz}, with a additional data sets included in the supplementary.
Figure \ref{fig:240wppmAzz} shows $A_{zz}$ as a function of filament strain for five stretching velocities.
Recall that for 240 wppm, the minimum velocity needed to achieve a high enough initial strain rate to observe true stress relaxation was 20 mm/s, and all lower velocities lead to erroneous relaxation times.
A comparison of $A_{zz}$ as a function of $\epsilon_D$ shows that the 5 mm/s experiment does not introduce significant chain stretch during the visco-capillary stretching regime (i.e. $\epsilon_D<1.8$).
Thus most of the stretching occurs in the Rayleigh-Plateau instability regime, and maximum stretch is only reached at very large $\epsilon_D$.
At 20 mm/s, we observe considerably more stretching during the visco-capillary stretching regime, and a much shorter strain required to reach maximum stretch.
At velocities above 20 mm/s, the chain stretch is almost a linear function of $\epsilon_D$ until it reaches the chain's maximum extensibility.

The effect of velocity on the chain stretch is more pronounced in Figure \ref{fig:240wppmAzz}(b), which shows the polymer stress as a function of filament strain.
Recall that for stress relaxation to be achieved, the polymer stress must exceed the capillary stress at small enough strain that the chain can undergo measurable stress relaxation.
At 5 mm/s, below the minimum initial $\dot{\epsilon}$, we observe that the polymer stress is always below the capillary stress until very high strains near the limitation of measurement (100 $\mu$m diameters).
Even then, there is very fast relaxation of the chain followed by no measurable stress relaxation.
At 20 mm/s, the minimum velocity to observe stress relaxation, we observe that the polymer stress exceeds the capillary stress at $\epsilon_D\approx5.2$, followed by a brief but observable stress relaxation of the chain.
At velocities above 20 mm/s, we see the polymer stress exceed the capillary stress at smaller $\epsilon_D$ followed by a very long stress relaxation period.
These results clearly demonstrate that the initial stretching of the chain in the visco-capillary stretching regime is fundamentally important in ensuring that the polymer stress exceeds the capillary stress at small enough $\epsilon_D$ that stress relaxation is observed.
The higher the velocity, the smaller the $\epsilon_D$ where $\sigma_p$ exceeds the capillary stress and the larger the stress relaxation window.

\begin{figure}[h!]
    \centering
    \includegraphics[width=0.95\textwidth]{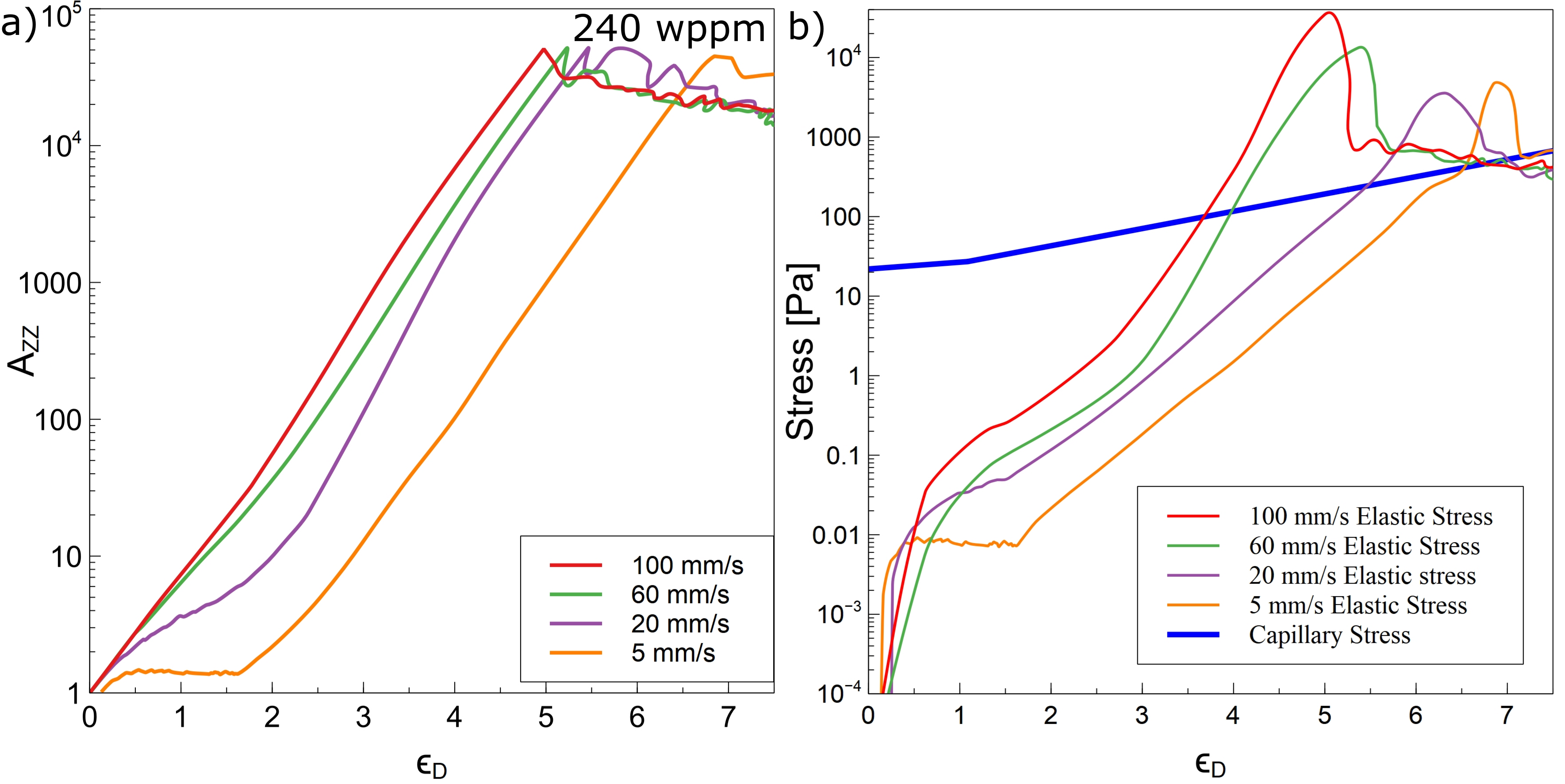}
    \caption{240 wppm, velocities 5 mm/s (initial $\dot{\epsilon}=2.5$ s$^{-1}$), 20 mm/s (initial $\dot{\epsilon}=10$ s$^{-1}$), 60 mm/s (initial $\dot{\epsilon}=26$ s$^{-1}$), 100 mm/s (initial $\dot{\epsilon}=37$ s$^{-1}$) a) Calculated Azz as a function of filament strain $\epsilon_D$ using experimental local strain-rate data b) polymer and capillary stress as a function of filament strain, $\epsilon_D$ using experimental local strain-rate data.}
    \label{fig:240wppmAzz}
\end{figure}

\begin{figure}[h!]
    \centering
    \includegraphics[width=0.95\textwidth]{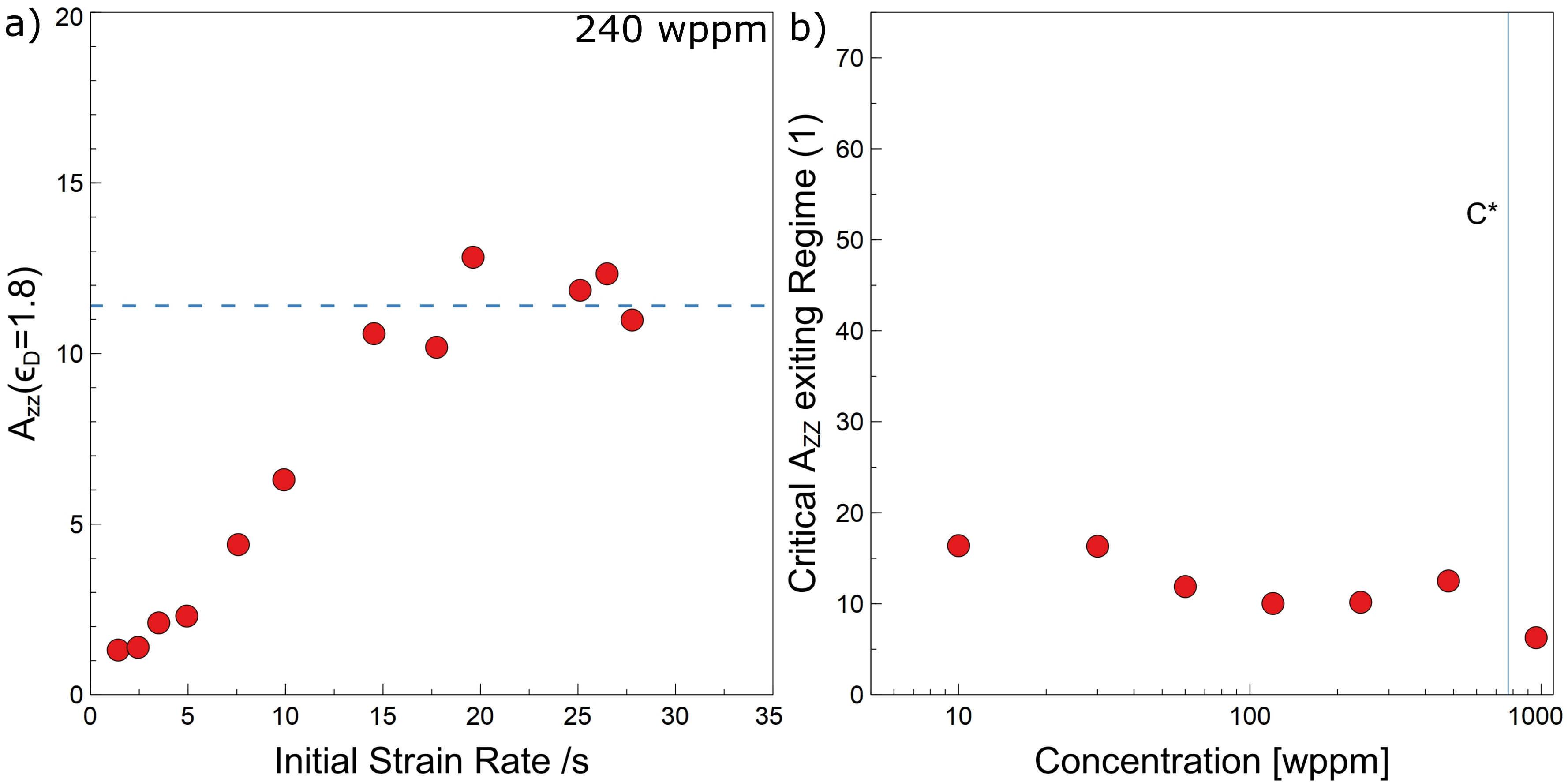}
    \caption{(a) $A_{zz}$ deteremined at $\epsilon_D$, the maximum strain in regime (1) versus the initial extension rate, (b) the asymptotic $A_{zz}$ as a function of polymer concentration.}
    \label{fig:MaxAzz}
\end{figure}

A more subtle point regarding the simulation results in Figure \ref{fig:240wppmAzz}(a-b) is that there is a minimum chain stretch required when exiting the visco-capillary stretching regime, i.e. stretching regime (1), to induce true stress relaxation.
The maximum induced strain in stretching regime (1) for a low-viscosity fluid is determined by its initial aspect ratio and volume.\cite{Barakat2021}
For the measurements in this study, the maximum strain is $\epsilon_D=1.8$.
An analysis of the $A_{zz}$ at $\epsilon_D=1.8$ gives a minimum chain stretch needed in regime (1) to achieve stress relaxation.
Figure \ref{fig:MaxAzz}(a) demonstrates there is more chain stretch with increasing initial strain rate (i.e. velocity)  until at high enough rates the chain stretch reaches an asymptotic value, $A_{zz}=11.5$.
This suggests that the at high enough rates, the chains are undergoing affine deformation.
Interestingly, this analysis was repeated for all concentrations and the asymptotic $A_{zz}$ in regime (1) is not a strong function of concentration.
Combining these results, we observe that there is a minimum chain stretch that must be achieved in regime (1) to observe stress relaxation, and this only occurs when the rate of deformation in regime (1) is greater than the inverse relaxation time.  
Furthermore, higher polymer concentrations, smaller relaxation times, reach a polymer stress greater than the capillary stress at earlier strains and therefore have a larger stress relaxation window.

\section{Additional advantages of the constant velocity scheme}

One major issue in standard CaBER measurements is the inertial recoil induced upon sudden stoppage of the top plate.
The recoil makes it difficult to capture the minimum thinning radius \cite{DeBlais2020a}.
One suggested solution to avoid recoil is the slow retraction method (SRM) \cite{campo2010}.
However, this method is clearly at odds with the results presented here, and most likely explains the very low relaxation times measured using this method for 500 wppm PEO (Mw=1$\times10^6$ g/mol) in water of $\lambda=$1.38 ms.
This is supported by relaxation time, $\lambda=2.4$ ms, measured using the free jet extensional rheometer measurement of the same concentration and molecular weight in water and glycol \cite{Christanti2002}.
Instead, our results here show that a constant plate separation velocity overcomes both recoil and ensures that the initial rate is large enough to ensure that the elasto-capillary thinning regime is true stress relaxation.
In fact, multiple velocities should be applied to determine whether a consistent relaxation time is observed. 

\section{Stretchability Index}

\begin{figure}[h!]
     \centering
     \includegraphics[width=\textwidth]{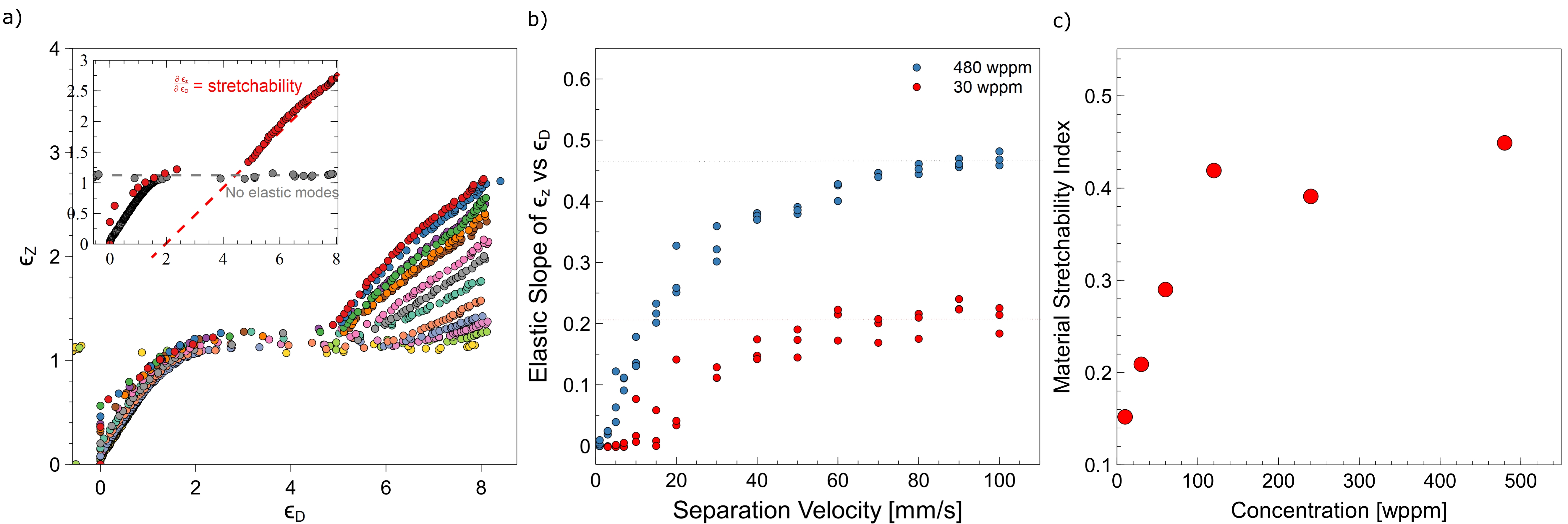} 
    \caption{(a) shows $\epsilon_z$ vs $\epsilon_D$ for 960 wppm at all velocities measured, the inset shows the definition of the stretchability factor determined from d$\epsilon_z$/d$\epsilon_D$ in the elastic stretching regime (b) stretchability factor as a function of plate separation velocity for 30 and 480 wppm, and (c) the asymptotic stretchability factor, referred to as the material stretchability index, as a function of concentration.}      \label{fig:Ez vs Ed}
\end{figure}

It is clear from users of capillary breakup indexers that the relaxation time, although a useful quantity, does not capture all the important physics of stretching.
For this reason, many users of capillary breakup measurements rely on shape and image analysis during the stretching process to compare different materials.
Such analysis is time consuming and not necessarily systematic.
In this work, we propose a different measure of the thinning behavior using the kinematic relationship between the radial Hencky strain, $\epsilon_D$, and the axial Hencky strain, $\epsilon_z$, measured during stretching.
The relationship between $\epsilon_D$ and $\epsilon_z$ is a function of the sample aspect ratio, the separation velocity, and the balance of stresses in the liquid bridge.
Figure \ref{fig:Ez vs Ed} (a) shows $\epsilon_z$ as a function of $\epsilon_D$ for a wide range of stretching velocities.
At low $\epsilon_D$, i.e. the visco-capillary regime $\epsilon_D<1.8$, there is almost no dependence of the kinematics on the stretching rate.
For $1.8<\epsilon_D<5$, we observe for all velocities a significant changes in $\epsilon_D$ with almost no change in $\epsilon_z$.
This region denotes the instability stretching regime, whereby the filament thins on its own.
Finally, for $\epsilon_D>5$, we observe a strong dependence of $\epsilon_z$ with stretching rate, which denotes the elasto-capillary stretching regime.
The kinematic relationship in this regime is an important indicator for how much a liquid bridge of this material can be stretched at a given velocity before the bridge breaks.

One measure of the materials stretchability is the slope of the kinematic curve in the elasto-capillary regime, see inset Figure \ref{fig:Ez vs Ed} (a).
Note that the stretchability factor is analagous to the inverse poisson ratio used in solid mechanics.
There are two important stretchability limits that have been determined experimentally and theoretically, namely viscous stretching, which has a slope of 2/3 and elastic stretching, which has a slope of 1.\cite{Marin2013}
Figure \ref{fig:Ez vs Ed} (b) shows the stretchability as a function of separation velocity for 30 and 480 wppm PEO solutions. 
For both concentrations, the stretchability of the material increases with increasing velocity and approach an asymptotic value.
The asymptotic value is a material parameter that is proportional to the polymeric stress induced in the material during the viscocapillary and instability stretching regimes.
Figure \ref{fig:Ez vs Ed} (c) shows the asymptotic stretchability, which we denote as the stretchability index, for all PEO concentrations tested, except 960 wppm.
Unlike the dilute concentrations, 960 wppm does not approach an asymptotic stretchability factor. 
Recall that C* for PEO solutions is 770 wppm.\cite{Graessley1980}
Exrapolation of the results shows that 960 wppm could exhibit fully elastic behavior (a slope of 1), at an estimated velocity of 139 mm/s.
This suggests that concentrations below the $C^*$ are distinct from the semi-dilute region, and not capable of fully elastic behavior.
For the dilute concentrations, the asymptotic stretchability factor increases significantly at low concentrations, but reaches a plateau of approximately 0.4 at 120 wppm.
Overall, the stretchability factor and index are unique material parameters, which define the degree of stretching for a given velocity and the overall shape of the material during stretching. 
We expect that these properties, in addition to the material relaxation time, will be very useful parameters in making correlations to process parameters.

\section{Conclusion}

This work highlights the importance of chain stretch in the visco-capillary stretching regime to achieve true stress relaxation in a capillary breakup measurement.
We showed that the stretching rate in the visco-capillary regime must be larger than 2/3 the inverse relaxation time to accurately measure the relaxation time of the material in the elasto-capillary thinning regime.
Below the minimum initial stretching rate, the relaxation time measured in the elasto-capillary regime is faster than the true polymer relaxation time, rate-dependent, and not a material parameter.
Comparison with literature revealed that many experimental studies reported much lower relaxation times for the same molecular weight polymer concentrations studied here.
This suggests that these studies did not reach the minimum visco-capillary stretching rate.
Furthermore, these results may explain the discrepancy in measured relaxation times using different capillary breakup measuring techniques.\cite{Mathues2018}
Unfortunately, the visco-capillary stretching regime is rarely reported in the literature and therefore it was not possible to confirm these hypotheses.

Another interesting result is that the strain required to achieve elasto-capillary thinning, $\epsilon_2$, is a strong function of the initial stretching rate and concentration, and was shown to correlate to the material relaxation time.
In other words, there is a minimum strain required on the filament to ensure that the correct relaxation time of the polymer is observed in the elaso-capillary thinning regime.
Lower polymer concentrations required a higher filament strain to observe true stress relaxation.
These results can be explained by understanding the relationship between filament strain and molecular strain.
At high enough initial stretching rates, the polymer chain should undergo affine deformation, and the polymer stress should increase with filament strain.
A higher filament strain required for lower concentrations indicates that there is a minimum chain stretch for a given concentration needed for the polymer stress to overcome the capillary stress.
These results have important implications for experiments.
For example, the capillary stress depends on the curvature of the filament, i.e. the diameter, and thus larger diameter filaments could be used to lower the required filament strain to observe true stress relaxation.
Coupling the experimental results with a single relaxation FENE model, we showed that there is indeed a minimum chain stretch required in the visco-capillary regime to achieve stress relaxation, which depends on the polymer concentration.

In many cases, the measure of relaxation time is not sufficient to correlate material properties to observed phenomena, and thus additional indexes are sought for material characterization.
One standard practice in capillary breakup experiments is to use a camera to observe the shape of the filament during breakup, and compare the filament shape as a function of time.
In this work, we show that a variable stretching rate, coupled with a direct measure of axial and radial filament strain allows for a kinematic description of the material, which is denoted as the stretchability factor.
More specifically, the stretchability factor is the slope of the kinematic curve in the elasto-capillary thinning regime, and is a direct measure of the materials stretchability.
The stretchability factor is a function of plate separation velocity, and approaches an asymptotic value at high enough stretching rates, i.e. faster than the inverse relaxation time.
The asymptotic value is denoted as the Material's Stretchability Index and is shown to be a function of polymer concentration.
At larger concentrations, below $C^*$, the Material Stretchability Index reaches an asymptotic value.
The stretchability index should prove a useful index in comparing materials for capillary breakup applications.

\section{Acknowledgements}
This work was supported by the National Science Foundation,
United States under grant no. CBET-1847140.

\bibliographystyle{bib_num} 
\bibliography{122023.bib}
\end{document}